\documentclass[preprint2]{aastex}

\shorttitle{Ultra-Compact Dwarfs}

\shortauthors{Phillipps et al.}

\begin{document}

\title{Ultra-Compact Dwarf Galaxies in the Fornax Cluster
\footnote{(To appear in {\em Astrophysical Journal})}}

\author{S. Phillipps}
\affil{Astrophysics Group, Department of Physics, University of Bristol,\\
Tyndall Avenue, Bristol BS8 1TL, U.K.}

\author{M.J. Drinkwater}
\affil{School of Physics, University of Melbourne, \\
Victoria 3010, Australia}

\author{M.D. Gregg}
\affil{Department of Physics, University of California, Davis, CA95616 and \\
Institute for Geophysics and Planetary Physics, Lawrence Livermore \\
National Laboratory, L-413, Livermore, CA 94550, USA}

\and

\author{J.B. Jones}
\affil{School of Physics and Astronomy, University of Nottingam\\
University Park, Nottingham NG7 2RD, U.K.}

\begin{abstract}
By utilising the large multi-plexing advantage of the 2dF spectrograph on
the Anglo-Australian Telescope, we have been able to obtain a complete
spectroscopic sample of all objects in a predefined magnitude range, 
$16.5 < b_j < 19.7$, regardless of morphology, in an area towards the
centre of the Fornax Cluster of galaxies. Among the unresolved or
marginally resolved targets
we have found five objects which are actually at the redshift
of the Fornax Cluster, i.e. they are extremely compact dwarf galaxies
or extremely large star clusters. All five have absorption line spectra.
With intrinsic sizes less than $1.1''$ HWHM
(corresponding to approximately
100pc at the distance of the cluster), they are 
more compact and significantly less luminous than other known
compact dwarf 
galaxies, yet much brighter than any globular cluster. 
In this letter we present new
ground based optical
observations of these enigmatic objects. In addition to
having extremely high central surface brightnesses, these objects show
no evidence of any surrounding low surface brightness envelopes
down to much fainter limits than is the case for, e.g.,
nucleated dwarf ellipticals. Thus, if they are not merely the
stripped remains of some other type of galaxy, then they appear to have
properties unlike any previously known type of stellar system. \end{abstract}


\clearpage 

\section{Introduction}

Almost since the beginning of extra-galactic astronomy it has commonly
been assumed that the full range of galaxy types and sizes is already 
known and catalogued, essentially a `what you see is what you get' (WYSIWYG)
point of view. Nevertheless, less conventional views have still been
espoused (e.g. Zwicky 1957, Arp 1963, Disney 1976). These might perhaps be
termed WYGIWYS, `what you get is what you see', as they posit that selection
effects have biased you against seeing anything else. Observation has indeed
provided new types of galaxy from time to time. Shapley's (1938) discovery
of extremely low luminosity dwarf spheroidals in the Local Group (see
also Shapley 1943), the
discovery by Zwicky (1957) himself of blue compact galaxies, and in more
recent years the finding of ever lower surface brightness galaxies
(Impey, Bothun \& Malin 1988), including some of remarkably large size
(Bothun et al. 1987), have all extended our view of the overall galaxy
population.

Even so, the currently standard view has galaxies, or stellar systems
in general, occupying discrete regions of the possible continuum of
size or luminosity versus surface brightness. Giant ellipticals have high
luminosities and high central surface brightnesses, the latter declining
somewhat towards the most luminous objects (Kormendy 1977). Dwarf ellipticals 
and dwarf spheroidals run from moderate luminosity and moderate surface 
brightness down to low values of both parameters (Ferguson \& Binggeli 
1994). Disc galaxies similarly have
lower surfaces brightnesses at lower luminosities (and later types),
running into irregular galaxies at the faint end (Binggeli, Sandage \&
Tammann 1984).
Finally globular
clusters have luminosities not much less than a dwarf galaxy, but
are very much smaller (more compact), so have very high central
surface brightnesses (Djorgovski 1995). The current situation can be 
summarised  schematically in a `Kormendy diagram' as
in figure 1 (following Ferguson \& Binggeli 1994). There are some known exceptions to the general rule. M32
is of significantly higher surface brightness than other dwarf (and
most giant) ellipticals, but may be either on the (very sparse)
extension of the
`normal' elliptical sequence to much lower luminosities or
a pathological case caused by
tidal stripping of a formerly larger system by the nearby M31 (Faber 1973). 
NGC 4486B
in Virgo may be another reasonably similar galaxy, but M32-like dwarfs
appear conspicuous by their absence in clusters such as Fornax
(Drinkwater \& Gregg 1998). Similarly Malin~1 is
an extreme low surface brightness disc galaxy of huge size, but even the
so called `Malin 1 cousins' are almost an order of magnitude smaller.

In the present paper we discuss the existence of what may be, if they
are not merely the remnants of some interaction, a new {\it class}
of galaxy, as opposed to exceptional cases. In the course of a complete 
spectroscopic survey of the Fornax Cluster (Drinkwater et al. 2000a = Paper I)
we have found several examples of extremely compact, yet moderately
luminous galaxies (Drinkwater et al. 2000b = Paper III;
Drinkwater et al. 2000c), which occupy a hitherto
empty region of the surface brightness -- size (luminosity) plane. We
discuss the spectroscopic discovery and the optical photometric properties
of these objects in the following sections 2 and 3, examining in
particular
whether they could be the tail (to high surface brightness
and/or low luminosity) of known galaxy types. Possible origins for
these systems are explored in section~4.

\section{The Fornax Cluster Spectroscopic Survey}

With the advent of modern multi-object spectroscopic facilities, exemplified
by the `two-degree field', or 2dF, spectrograph on the Anglo-Australian
Telescope (Taylor, Cannon \& Parker 1998), entirely new ways of
surveying the universe have become possible. We have taken advantage of
these opportunities by carrying out the first deep, all-object survey
of a galaxy cluster region. That is, we have obtained spectra for all
the objects in the region between set magnitude limits, regardless of
apparent morphology (i.e. `star' or `galaxy'). One of the main reasons
behind this strategy was to test the hypothesis that compact, high
surface brightness galaxies have been ignored in conventional galaxy
surveys because of their small isophotal size or, indeed, because they
are indistinguishable from stars in the ordinary ground-based imaging
which provides the input catalogues for galaxy spectroscopic surveys
(e.g. Disney \& Phillipps 1983).

The system we use, 2dF on the AAT, has two sets of 200 fibres, each
feeding a separate spectrograph and allowing the simultaneous observation
of 400 objects. In the central 2 degree diameter area of the Fornax
Cluster there are around 4000 objects in the chosen magnitude range 
$16.5 \leq b_j \leq 19.7$ for galaxies, $16.5 \leq b_j \leq 20.0$
for `stars', thus requiring at least 10 separate observations.
Of course, even towards the cluster centre, the majority of objects are
not cluster galaxies, the numbers being dominated by foreground Galactic
stars and background field galaxies (Paper I). We used the 300B grating,
giving wavelength resolution of approximately 9 \AA $\;$ (4.3 \AA $\;$ per pixel)
over the range 3600 - 8100 \AA, the same set-up as for the general 2dF Galaxy
Redshift Survey (e.g. Folkes et al. 1999). Total integration
times (the observations are usually subdivided to assist with cosmic ray
removal) ranged from 1 hour for the brighter, higher surface brightness
objects to about 3 hours for the fainter low surface brightness galaxies.

The overall input catalogue for the survey is derived from Automated
Plate Measuring (APM) machine scans of UK Schmidt Telescope plates of
the area (see Irwin, Maddox \& McMahon 1994). The APM catalogue lists
positions, magnitudes and morphological classifications (`star' = unresolved,
`galaxy' = resolved, or `merged' = overlapping images, usually of a star
and a fainter galaxy). The APM magnitudes of unresolved images are
internally calibrated, but we also checked them using our own CCD photometry
from the CTIO Curtis Schmidt (see Paper I), resulting in a small zero point
adjustment compared to the default APM magnitudes. Magnitudes
for resolved galaxy images were derived
as discussed in Paper I. We then chose to target spectroscopically
all objects with magnitudes $16.5 \leq b_j \leq 19.7$, (or 20.0 for stars)
where $b_j$ is the natural
UKST photographic B band defined by the IIIaJ emulsion and GG395 filter (see
e.g. Blair \& Gilmore 1982).

The spectra were reduced using both DOFIBERS within IRAF
\footnote{IRAF is distributed by the National Optical Astronomy
Observatories, operated by AURA, Inc.,
under cooperative agreement with the National Science Foundation.}
 and the instrument
specific 2dFDR software, with essentially
identical results. No attempt is made to `flux' the spectra. Further
particulars, especially on the sky subtraction, are given in Paper I
and we do not repeat them here.

Once we have reduced spectra, we determine redshifts and approximate spectral
types uniformly for all objects in the survey via the cross-correlation
method (Tonry \& Davies 1979). All the object spectra (irrespective
of image classification) are compared with a set of standard templates;
nine stellar
templates for types A3V through M5V from Jacoby, Hunter \& Christian (1984)
plus emission line galaxy and QSO templates. Note that the stellar
templates result in equally good correlations for absorption line galaxies
as would actual galaxy templates (see Paper I).
Cross-correlations are calculated using RVSAO (Kurtz \& Mink 1998), which
determines a redshift ($cz$), its error, and the Tonry-Davis $R$
coefficient which measures the significance of the match. We accept only
identifications with $R > 3$ and in addition check by eye for any possible
misidentifications. The rms velocity error found from repeat observations
is $\simeq 64$ km~s$^{-1}$, consistent with the values reported by RVSAO
and with external comparisons (primarily with Hilker et al. 1999b).

As of the end of 1999, we had observed 92\% of our targets in the desired
magnitude range in a 2 degree diameter field in the centre of the
cluster (centred close to NGC 1399)
and successfully obtained redshifts for 94\% of those.
The results show the Fornax Cluster to be well
separated from the rest of the field in redshift space. From our 2dF
results alone (which correspond to dwarf galaxies with $-15.0 \leq M_B
\leq -11.5$) we find a cluster redshift $cz_{mean} = 1450 \pm 70$ km~s$^{-1}$
and a velocity dispersion $\sigma = 380 \pm 50$ km~s$^{-1}$, in good agreement
with the values for brighter galaxies (e.g. Jones \& Jones 1980). There
are no galaxies with $cz < 900$ or between 2300 and 3000 km~s$^{-1}$.

\section{Cluster Compact Galaxies}

From the reduced spectra it is clear that a small but not insignificant
fraction of the `stars' actually have galaxy spectra with recession
velocities greater than 1000 km~s$^{-1}$. A number of these are fairly
distant compact emission line galaxies
(see Drinkwater et al. 1999 = Paper II), while a few are similarly distant
compact galaxies but with absorption line spectra .
In addition, as reported in Paper III,
four unresolved objects and a fifth marginally resolved object 
turn out to have velocities
clearly indicating membership of the Fornax Cluster. Even without further
analysis, their lack of obvious resolution on UKST survey plates already
marks them down as unusual objects, since at an 
assumed Fornax distance of 20 Mpc
(distance modulus 31.5; Drinkwater, Gregg \& Colless 2001), 
a scale-length of
$1''$, say, would correspond to just 100pc. The five objects -- to which we attach the provisional classification ultra-compact 
dwarf or UCD -- are listed
in Table 1. Images/finding charts are provided in Paper~III.

Note that the existence of these five UCDs increases the known number
of cluster dwarfs in the relevant magnitude range by about 10\% if we take
the number of `certain' members seen in our area
at slightly brighter magnitudes in Ferguson's 
(1989) Fornax Cluster Catalog (FCC) and extrapolate using
Ferguson \& Sandage's (1988 = FS) cluster dwarf
luminosity function. (Our own 
sample is too limited by the surface brightness limit for successful
spectroscopic observations to define total numbers). The UCDs 
themselves are {\it not} in the FCC (even as background galaxies),
presumably because they are still unresolved even on the much higher
resolution Las Campanas plates used by FS. We also note, for later
reference, that {\em all} the dwarf cluster members which were
identified by
FS are clearly visible (and well resolved) on our UKST plate and film
material.

The two brightest of the ultra-compacts (UCD3 and UCD4)
turn out to have been observed independently by Hilker et al. (1999b; see table 1),
who obtained spectra for some 50 objects down to $V = 20$ very close to the 
central cluster galaxy NGC~1399. They found them to be slightly resolved
in their imaging data, and we see similar resolution of UCD3 in our
CTIO CCD imaging and in $R$ band Tech Pan photographic films from the UKST (which have slightly better resolution than the $J$ plates used in
the APM catalogue).
The other four ultra-compact objects appear entirely stellar in
all our ground based images, see figure 2. 

The FWHM of the seeing on the Tech Pan data is about $2.3''$ while UCD3, 
the largest object, has an image FWHM around $3.2''$. A very simple
deconvolution then suggests an intrinsic radius (HWHM) of about $1.1''$
or 110pc.
Obviously the other UCDs must be smaller still. Note that at
the distance of Fornax, even the physically smallest Local Group dwarf
spheroidal, Leo II ($M_V \simeq -10.1$, effective radius 180pc;
Mateo 1998),
would be reasonably well resolved with an intrinsic
half-light radius of $2''$. The one relatively
high surface brightness Local Group dwarf spheroidal, Leo I 
($M_{V} \simeq -11.5$), would be even better resolved with effective
radius around $3''$ (van den Bergh 1999).

The UCDs have absolute magnitudes within the range
$-14.0 < M_B < -11.5$, placing them in the lower range of luminosities 
for known dwarf systems (Mateo 1998) as shown in figure 3.
However, previously known low luminosity dwarfs have low surface brightnesses
too, so are morphologically quite distinct from the present group of galaxies
(see figure 1, where the UCDs are shown as the upward pointing
arrows, since we have only lower limits to their true unconvolved 
central surface
brightnesses). Put the other way round, our objects are much fainter than
previously discovered nearby compact (high surface brightness) galaxies (e.g.
those of Drinkwater \& Gregg 1998; see also Drinkwater \& Hardy 1991 for 
blue compact dwarfs). They are also very much fainter 
and smaller than 
the compact star-forming dwarfs seen at intermediate to high redshift
(e.g. Guzman et al. 1998, Paper II) and do not appear to have any direct
relationship to them. 

On the other hand, the
UCDs are much brighter than any Galactic globular clusters or
known globulars around NGC~1399 (Bridges, Hanes \& Harris 1991). The slightly
resolved UCD3 is, of course, also considerably larger than any
known globular (which have
half light radii up to about 10pc; Djorgovski 1995). 
A third class of small stellar system which might be
comparable to our objects are the M32-like dwarf ellipticals, but none
of the candidates for such a galaxy listed in the FCC
has yet been found to be a cluster member (Drinkwater \& Gregg 1998). Finally,
there are the nuclei of nucleated dwarf ellipticals. These dE,N nuclei
do span the luminosity range of our ultra-compacts (Binggeli \& Cameron 1991), see figure 3,
and are, of course, also very small. However, our UCDs can not
be the nuclei of `ordinary' dE,N 
as we would be able to see the surrounding dE itself (recall that we
can detect and resolve all the FS dEs and dE,Ns in our area). 

To examine whether our objects could be nuclei of very low surface
brightness dEs, we have further analysed our 
Tech Pan $R$ images of the five UCDs, 
and also new AAT prime focus CCD images ($0.23''$ pixels)
for the two brightest objects (20 minute $B$ band
and 10 minute $V$ band exposures obtained in service time
with approximately $1.4''$ seeing), 
to look for any remaining diffuse surrounding starlight.
None is detected in the CCD images
down to 3$\sigma$ surface brightness limits
around $26.6 V$ mags arcsec$^{-2}$ on scales of a few 
hundred parsecs. Similarly for the deep Tech Pan imaging,
we see no sign of any surrounding envelope for any of the UCDs. (We do confirm that UCD3 is resolved, as previously suspected, but even
there see no measurable light outside the central core).
For UCD1, the faintness of the core and the absence of
nearby objects allows us to determine an even lower limit of $27.1 R$ 
mags arcsec$^{-2}$ on similar scales (the $4''$ to $6''$ range
in the profiles shown in figure 2). 

These limits are very much lower than seen for known dE,N galaxies.
In particular, if we carry out exactly the same experiment for {\em all}
59 of the known FS dE,N galaxies in the area covered by
our photometry, down to the lowest
known luminosity objects of this type, we can easily measure the mean  
surface
brightness for each. We find that they all have surface
brightnesses 1 to 4 magnitudes higher than our 3 sigma upper limits on
any surrounding galaxy to our UCDs. Thus if our objects are the
nuclei of dEs, then they are dEs of unprecedented faintness, 
well separated from the known population, {\em or}
the surrounding galaxy is no longer present, for example having been
tidally stripped (see section 4).

Spectroscopically the UCDs look similar to ordinary early
type dwarfs in the FCSS sample (see figure 3 of paper III), with no
discernable emission lines. The template matching suggests
slight differences, with K-type spectra consistent with old metal rich stellar populations,
for the ultra-compacts, compared to generally younger F or G-type populations
in dEs and dE,Ns. This may argue for stellar populations in the ultra-compacts
rather like those in giant ellipticals in the cluster, which appear to be
uniformly old and of high metallicity (Kuntschner \& Davies 1998).  
Both metal rich globular clusters and M32 also have this type of spectrum
(Hilker et al. 1999b). (Note that Hilker et al. suggest that their CGF5-4
(our UCD4) may differ from this (specifically from CGF1-4 = UCD3), in having a 
spectrum more like a metal poor globular or a dE nucleus, but we see no 
significant difference in our 2dF spectra).

\section{Discussion and Summary}

As intimated in the introduction, a potentially major shortcoming
of morphology based catalogues of dwarf galaxies in clusters has been
the possibility that only subsets of cluster members with familiar
properties are selected. Specifically, high surface brightness dwarf
galaxies may be mistaken for background giants, or indeed be so
compact as to look like stars. Our all-object spectroscopic survey
has shown that this is indeed the case. Very compact, small scale size, 
high surface brightness dwarfs do exist in clusters. The less extreme
objects (Drinkwater \& Gregg 1998; Drinkwater et al. 2000c) are probably
an extension of classically identified dwarfs to rather higher surface
brightness at a given magnitude, thus blurring somewhat the
surface brightness - luminosity relation. We discuss the significance
of these elsewhere. However, the group of objects identified by
Drinkwater et al. (2000b,c), which were previously
confused with stars, appear to be disjoint from any other known type
of stellar system in the surface brightness - luminosity plane (figure 1).

Several possibilities as to their nature suggest themselves. They
may be genuine (i.e. primordial) high central density galaxies, that
is a new class of stellar system not previously identified; they may
be super-massive versions of globular clusters; they may be the
nuclei of extremely low surface brightness dE galaxies; or they may be
tidally distorted remnants of normal dwarfs, either small M32-like
objects or the nuclei of former nucleated dwarfs or late type spirals.

Relatively little can be said for or against the notion that the
UCDs are smaller versions of M32, since the evolution of M32
itself is not clear (see van den Bergh 1999). We can note the lack of actual
(i.e. moderate luminosity) M32 analogues in Fornax (Drinkwater \& Gregg
1998), but of course some sort of post-formation tidal limiting effect
may be more damaging to less massive systems. Similarly it is impossible
to discount totally the possibility that the UCDs are really the
nuclei of larger galaxies, though in this case their hosts would have
to have surface brightnesses far below those of known dE galaxies
(and so would still represent a class of galaxy disjoint in their
properties from those already known).

If the UCDs are produced in the original galaxy formation
process, it is possible that they are some sort of super-massive star
cluster, perhaps a kind of globular cluster, rather than a `real' galaxy
(though one might argue that this is a merely semantic distinction).
NGC~1399 is well known to have a large population of globulars (Grillmair 
et al. 1994). The UCDs are all situated
within $30'$ (150 kpc) of NGC~1399, though of course our survey field
has only twice this radius and some of the ultra-compacts are actually
nearer to other large galaxies, in projection, than to NGC~1399. 
(UCD3 is very close to NGC~1404 for example). Kissler-Pattig et al.
(1999) have suggested that some of the globulars follow the 
dynamics of the cluster as a whole, rather than the halo of NGC~1399
itself, and that these may have been tidally removed from other cluster
galaxies (see also West et al. 1995). Alternatively, West et al. also 
consider the possibility that intra-cluster globulars might form {\it in situ}.
Of course, our objects are a factor 10 more luminous than any known
globulars, so they would have to be massive `super-globulars' if they
were associated with such a population
(see also Goudfrooij et al. 2000). Note that the radial distribution
of the UCDs is consistent with that suggested by West et al.
for an intra-cluster population but significantly more dispersed than the
NGC~1399 globular cluster system discussed by Grillmair et al. 
(see Paper~III).

Bassino et al. (1994) have discussed whether some, at least, of 
the globular clusters in rich systems, such as that
of NGC~1399, could be the remnant nuclei of former nucleated dwarf
ellipticals which have been accreted by the central cluster galaxy
(NGC~1399 has cD galaxy like properties). They also note that remnants an order
of magnitude more massive than normal globulars should also be formed
by this process. This would be consistent with the luminosities of
our objects as shown in figure 3. This possibility has recently been 
considered in more detail by Bekki, Couch \& Drinkwater (2001).
Alternatively, it may be possible that similar remnants can form from
the `shredding' by tidal forces of late type spirals with small nuclear 
bulges (Moore et al. 1996), since galaxies like M33 appear to have
`bulges' more akin to a giant star cluster (e.g. Mighell \& Rich 1995).

Simulations suggest that in either version of this remnant picture we
might still expect to see surrounding very low surface brightness halos
of stars, at least for several Gyr, but as noted above, none
have yet been detected down to quite faint limits. On the other hand, the
disrupted material, including surviving nuclei, would be expected to
be more concentrated to the cluster centre than the galaxies as
a whole (White 1987), as is observed (see Paper III).

Finally there is the possibility that the UCDs represent  a genuinely
new class of galaxy.
In the cold dark matter (CDM) picture of galaxy formation (e.g. Kauffmann,
Nusser \& Steinmetz 1997), small dense halos should collapse at high redshifts and
subsequently merge into larger structures. Recent simulations
(e.g. Moore et al. 1998) have the resolution to begin to see the
details of the evolution of small halos within larger cluster sized
structures. However, the available resulution still limits investigation
to masses $\sim 10^{9}  M_{\odot}$ and we do not yet know the lower
mass limit for individual halos in a cluster environment, nor indeed,
the behavior of the baryonic (visible galaxy) component within these
sub-structures. Establishing some limits on low mass but dense galaxies
could provide interesting future constraints on these models, providing
the systems we are seeing are (within) primordial halos. Previously
known very low mass galaxies are of low surface brightness (low
visible baryonic surface density) and probably dark matter dominated
throughout (Carignan \& Freeman 1988; Mateo 1998), so high
surface brightness dwarfs may provide interesting counter-examples. (It
may be hard to see, for example,  why they are so compact optically if 
they have dark matter core radii of several hundred parsecs like `normal'
dwarfs, or if the central dark matter density is similar for all dwarfs).

In summary, the observations to date do seem to suggest that we are
seeing a new
type of stellar system. The UCDs are certainly much more
luminous than normal globular clusters but are at the same time much
less luminous than known compact dwarf galaxies. They do have luminosities
similar to known faint dwarfs (e.g. in the Local Group), but have
entirely different morphologies, previously known extreme dwarfs all having
low surface brightnesses. There remains the possibility that they are the
remnants of some more well known type of galaxy, after disruption in the
potential of the cluster. In particular they share a commom luminosity range 
with the nuclei of nucleated dwarf ellipticals,
though no surrounding `host' or the remains thereof is visible. 
Further observations at
much higher resolution, both spatial and spectroscopic, are required in
order to elucidate the true nature of these enigmatic objects.
HST images and VLT spectra have recently been obtained for this purpose.

\acknowledgements
The FCSS would not have been possible without
the development of 2dF at the Anglo-Australian Observatory. 
We would also like to thank the SuperCOSMOS unit at the
Royal Observatory Edinburgh for scanning the UK Schmidt films
used, PATT and ATAC for the award of telescope time for
our long term FCSS project,
our colleagues Jon Davies, Julia Deady, Quentin Parker, Elaine
Sadler and Rodney Smith for their part in the overall FCSS,
Harry Ferguson,
Michael Hilker and Ben Moore for useful discussions and
Sidney van den Bergh for encouraging us to consider the UCDs
as a genuinely new class of galaxy.

MJD acknowledges support from a Australian Research Council Large Grant,
and JBJ the support of the UK Particle Physics and
Astronomy Research Council. 
Part of the work reported here was done at the Institute of Geophysics
and Planetary Physics, under the auspices of the U.S. Department of
Energy by Lawrence Livermore National Laboratory under contract No. 
W-7405-Eng-48. It is also based upon work supported by the National
Science Foundation under grant No. 9970884.


\clearpage

{
\epsscale{0.99} \plotone{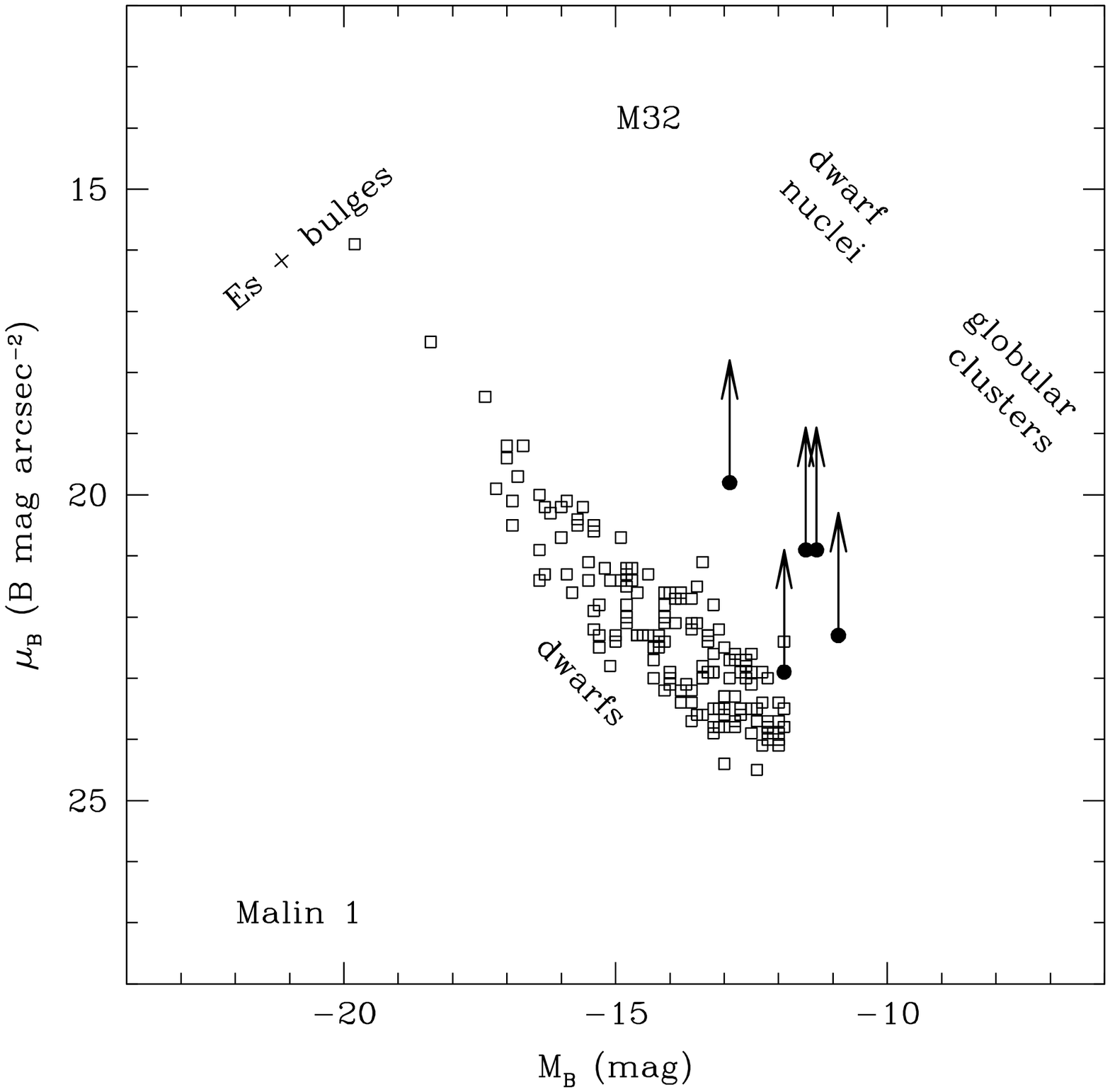} 
\figcaption{The absolute magnitude -- central surface brightness plane 
for stellar
systems and subsystems. The squares indicate our measurements of dwarf
galaxies in the Fornax Cluster and the filled circles the new Fornax
compact objects (the surface brightness estimates are lower
limits).  The positions of other populations are from Ferguson \&
Binggeli (1994).}
}

{\vspace{-0.5cm}
\epsscale{0.99} \plotone{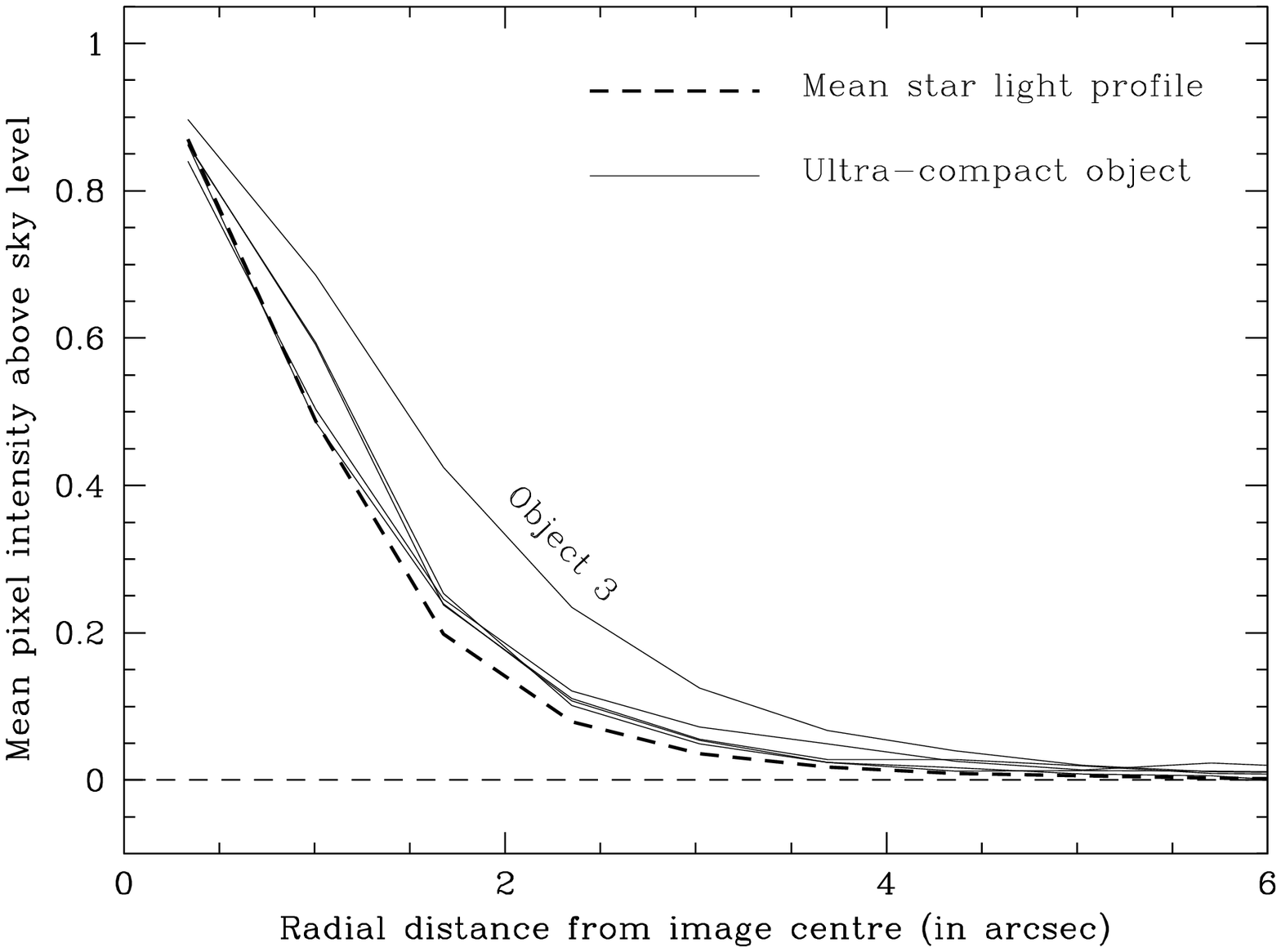}
\figcaption{Radial R band intensity profiles (normalised to unity at
the centre) for the five UCDs (thin lines) as derived from SuperCOSMOS 
scans of UKST
Tech Pan films. For comparison the thick
dashed line shows
the average profile of neighbouring stars of similar magnitude.}
{

{
\epsscale{0.99} \plotone{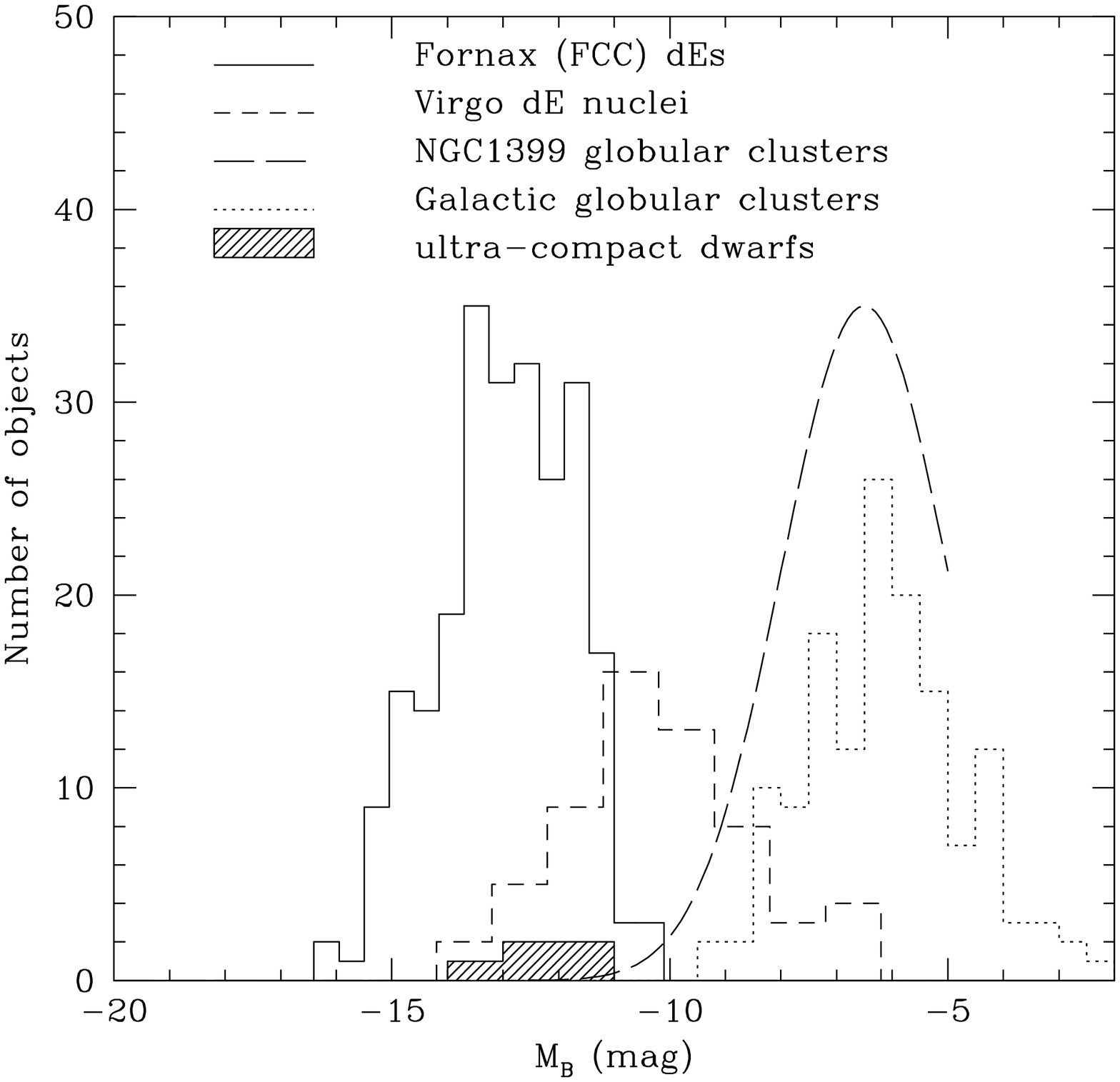}
\figcaption{The distribution of absolute magnitude of the compact objects
(filled histogram) compared to dEs in the Fornax Cluster (FCC;
solid histogram), nuclei of dE,Ns in the Virgo Cluster
(Binggeli \& Cameron 1991; short dashes), a model fit to the globular
clusters around NGC 1399 (Bridges et al.\ 1991; long dashes)
and Galactic globular clusters (Harris 1996; dotted). The
magnitude limit of our survey corresponds to $M_B=-11.5$.}
}

\clearpage

\begin{table*}
\begin{center}
\begin{tabular}{lccllcl}
\hline
Name& IAU Name & RA (J2000) Dec & $b_j$& $M_B$ & cz & Hilker\\
   &  &                & (mag) & (mag) & (kms$^{-1}$) & Name\\
\hline
UCD1&FCSS J033703.3-353804 & 03 37 03.30 $-$35 38 04.6& 19.85& $-11.6$& 1507 \\
UCD2&FCSS J033806.3-352858 & 03 38 06.33 $-$35 28 58.8& 18.85& $-12.6$& 1328 \\
UCD3&FCSS J033854.1-353333 & 03 38 54.10 $-$35 33 33.6& 17.68& $-13.8$& 1595& CGF1-4\\
UCD4&FCSS J033935.9-352824 & 03 39 35.95 $-$35 28 24.5& 18.82& $-12.7$& 1936& CGF5-4\\
UCD5&FCSS J033952.5-350424 & 03 39 52.58 $-$35 04 24.1& 19.66& $-11.8$& 1337 \\
\hline
\end{tabular}
\end{center}
\caption{The Ultra-Compact Dwarfs  \label{tbl-1}}
\end{table*}


\begin{thebibliography}{}
\bibitem[Arp 1963]{A63}
Arp H.C., 1963, ApJ, 142, 402
\bibitem[Bassino et al. 1994]{B94}
Bassino L.P., Muzzio J.C., Rabolli M, 1994, ApJ, 431, 634
\bibitem[Bekki et al. 2001]{BCD01}
Bekki K., Couch W.J, Drinkwater M.J., 2001, ApJ, 552, L105
\bibitem[Binggeli and Cameron 1991]{BC91}
Binggeli B., Cameron L.M., 1991, A\&A, 252, 27
\bibitem[Binggeli et al. 1984]{BST}
Binggeli B., Sandage A., Tammann G.A., 1984, AJ, 90, 1681
\bibitem[Blair and Gilmore 1982]{BG82}
Blair M., Gilmore G., 1982, PASP, 94, 742
\bibitem[Bothun et al. 1987]{BIMM}
Bothun G., Impey C., Malin D.F., Mould J., 1987, AJ, 94, 23
\bibitem[Bridges et al. 1991]{B91}
Bridges T.J., Hanes D.A., Harris W.E., 1991, AJ, 101, 469
\bibitem[Carignan and Freeman 1988]{CF88}
Carignan C., Freeman K.C., 1988, ApJ, 332, L33
\bibitem[Disney 1976]{D76}
Disney M.J., 1976, Nat, 263, 573
\bibitem[Disney and Phillipps 1983]{DP83}
Disney M.J., Phillipps S., 1983, MNRAS, 205, 1253
\bibitem[Djorgovski 1995]{D95}
Djorgovski S., 1995, ApJ, 438, L29
\bibitem[Drinkwater and Gregg 1998]{DG98}
Drinkwater M.J., Gregg M.D., 1998, MNRAS, 296, L15
\bibitem[Drinkwater and Hardy 1991]{DH91}
Drinkwater M.J., Hardy E., 1991, AJ, 101, 94
\bibitem[Drinkwater et al. 2001]{DGC01}
Drinkwater M.J., Gregg M.D., Colless M.M., 2001, ApJ, 548, L139
\bibitem[Drinkwater et al. 1999]{D99}
Drinkwater M.J., Phillipps S., Gregg M.D, Parker Q.A., Smith R.M., Davies J.I.,
 Jones J.B., Sadler E.M., 1999, ApJ, 511, L97 (Paper II)
\bibitem[Drinkwater et al. 2000a]{D00a}
Drinkwater M.J., Phillipps S., Jones J.B., Gregg M.D., Deady J.H., Davies J.I.,
 Parker Q.A., Sadler E.M., Smith R.M., 2000a, A\&A, 355, 900 (Paper I) 
\bibitem[Drinkwater et al. 2000b]{D00b}
Drinkwater M.J., Jones J.B., Gregg M.D., Phillipps S., 2000b, PASA, 17, 227 (Paper III)
\bibitem[Drinkwater et al. 2000c]{D00c}
Drinkwater M.J., et al., 2000c, in Kraan-Korteweg R.C., Henning P.A., Andernach 
H., eds, Mapping the Hidden Universe, in press
\bibitem[Faber 1973]{F73}
Faber S.M., 1973, ApJ, 179, 731
\bibitem[Ferguson 1989]{FCC}
Ferguson H.C., 1989, AJ, 98, 367 (FCC)
\bibitem[Ferguson and Sandage 1988]{FS88}
Ferguson H.C., Sandage A., 1988, AJ, 96, 1520 (FS)
\bibitem[Ferguson and Binggeli 1994]{FB94}
Ferguson H.C., Binggeli B., 1994, A\&ARv, 6, 67
\bibitem[Folkes et al. 1999]{Fo99}
Folkes S., et al., 1999, MNRAS, 308, 459
\bibitem[Goudfrooij et al. 2000]{G00}
Goudfrooij P., Mack J., Kissler-Pattig M., Meylan G, Minniti D., 2001, MNRAS, 322, 643
\bibitem[Grillmair et al. 1994]{G94}
Grillmair C.J., et al., 1994, ApJ, 422, L9
\bibitem[Guzman et al. 1998]{G98}
Guzman R., Jangren A., Koo D.C., Bershady M.A., Simard L., 1998, APJ, 495, L13
\bibitem[Harris 1996]{H96}
Harris W.E., 1996, AJ, 112, 487
\bibitem[Hilker et al. 1998]{H99a}
Hilker M., Kissler-Patig M., Richtler T., Infante L., Quintana H., 1999a, A\&AS,
134, 59
\bibitem[Hilker et al. 1999]{H99b}
Hilker M., Infante L., Vieira G., Kissler-Pattig M., Richtler T., 1999b, A\&AS,
134, 75
\bibitem[Impey et al. 1988]{IBM}
Impey C.D., Bothun G.D., Malin D.F., 1988, ApJ, 330, 634
\bibitem[Irwin et al. 1994]{I94}
Irwin M.J., Maddox S.J., McMahon R., 1994, Spectrum, 2, 14
\bibitem[Jacoby et al. 1984]{J84}
Jacoby G.H., Hunter D.A., Christian C.A., 1984, ApJSup, 56, 257
\bibitem[Jones and Jones 1980]{JJ80}
Jones B.J.T., Jones J.E., 1980, MNRAS, 191, 685
\bibitem[Kauffmann et al. 1997]{K97}
Kauffmann G., Nusser A., Steinmetz M., 1997, MNRAS, 286, 795
\bibitem[Kissler-Pattig et al. 1999]{K99}
Kissler-Pattig M., Grillmair C.J., Meylan G., Brodie J.P., Minniti D., 
Goudfrooij P., 1999, AJ, 117, 1206
\bibitem[Kormendy 1977]{K77}
Kormendy J.C., 1977, ApJ, 104, 340
\bibitem[Kuntschner and Davies 1997]{KD98}
Kuntschner H., Davies R.L., 1998, MNRAS, 295, L29
\bibitem[Kurtz and Mink 1998]{KM98}
Kurtz M.J., Mink D.J., 1998, PASP, 110, 934
\bibitem[Mateo 1998]{M98}
Mateo M., 1998, ARA\&A, 36, 435
\bibitem[Mighell and Rich 1995]{MR95}
Mighell K.J., Rich R.M., 1995, AJ, 110, 1649
\bibitem[Moore et al. 1996]{M+96}
Moore B., Katz N., Lake G., Dressler A., Oemler A., 1996, Nat, 379, 613
\bibitem[Moore et al. 1998]{M+98}
Moore B., Governato F., Quinn T., Stadel J., Lake G., 1998, ApJ, 499, L5
\bibitem[Shapley 1938]{S38}
Shapley H., 1938, Harv. Bull., 908, 1
\bibitem[Shapley 1943]{S43}
Shapley H., 1943, Galaxies, Blakiston, Philadelphia
\bibitem[Taylor et al. 1998]{T98}
Taylor K., Cannon R.D., Parker Q.A., 1998, in McLean B.J., Golombek D.A.,
Hayes J.J.E., Payne H.E., eds, New Horizons from Multi-Wavelength Sky Surveys,
p.135
\bibitem[Tonry and Davies 1979]{TD79}
Tonry J.L., Davis M., 1979, AJ, 84, 1511
\bibitem[van den Bergh 1999]{vdB99}
van den Bergh S., 1999, The Galaxies of the Local Group, Cambridge U.P.
\bibitem[West et al. 1995]{W95}
West M.J., C\^{o}t\'{e} P., Jones C., Forman W., Marzke R.O., 1995, ApJ, 453, L77
\bibitem[White 1987]{W87}
White S.D.M., 1987, MNRAS, 227, 185
\bibitem[Zwicky 1957]{Z57}
Zwicky F., 1957, Morphological Astronomy, Springer Verlag, Berlin
\end{thebibliography}
\end{document}